\documentclass[preprint,number,sort&compress,twocolumn,3p]{elsarticle}
\pdfoutput=1
\usepackage{rotating}
\usepackage{graphicx}
\usepackage{array}
\usepackage{amsmath}
\usepackage{color}
\usepackage{amssymb}
\usepackage{slashed}
\usepackage[pdftex,hyperfootnotes=false,pdfpagelabels]{hyperref}

\begin{document}

\title{Tricking Landau-Yang: How to obtain the diphoton excess from a vector resonance\tnoteref{t1}}

\tnotetext[t1]{This article is registered under preprint number: DESY 15-256, arXiv:1512.06833}

\author{Mikael Chala}
\author{Michael Duerr}
\author{Felix Kahlhoefer\fnref{fn1}}
\author{Kai Schmidt-Hoberg}

\fntext[fn1]{felix.kahlhoefer@desy.de}

\address{DESY, Notkestra\ss e 85, D-22607 Hamburg, Germany}

\begin{abstract}
We show that contrary to naive expectations the recently observed diphoton excess can be explained by a vector resonance,
which decays to a photon and a light scalar $s$, followed by a decay of the scalar into two photons: $Z' \rightarrow \gamma s \rightarrow 3 \gamma$.
As the two photons from the scalar decay are highly boosted, the experimental signature is an apparent diphoton final state. 
In fact all the necessary ingredients are naturally present in $Z'$ models: Additional fermions with electroweak quantum numbers are
required in order to render the theory anomaly free and naturally induce the required effective couplings, while the hidden Higgs which gives mass to the $Z'$ can be very light. In particular no new coloured states are required in this framework.
We also show that in such a setup the width of the resonance can be rather large, while all couplings remain perturbative.
\end{abstract}


\maketitle

\section{Introduction}

\begin{figure}[tb]
\centering
\includegraphics[height=0.15\textheight]{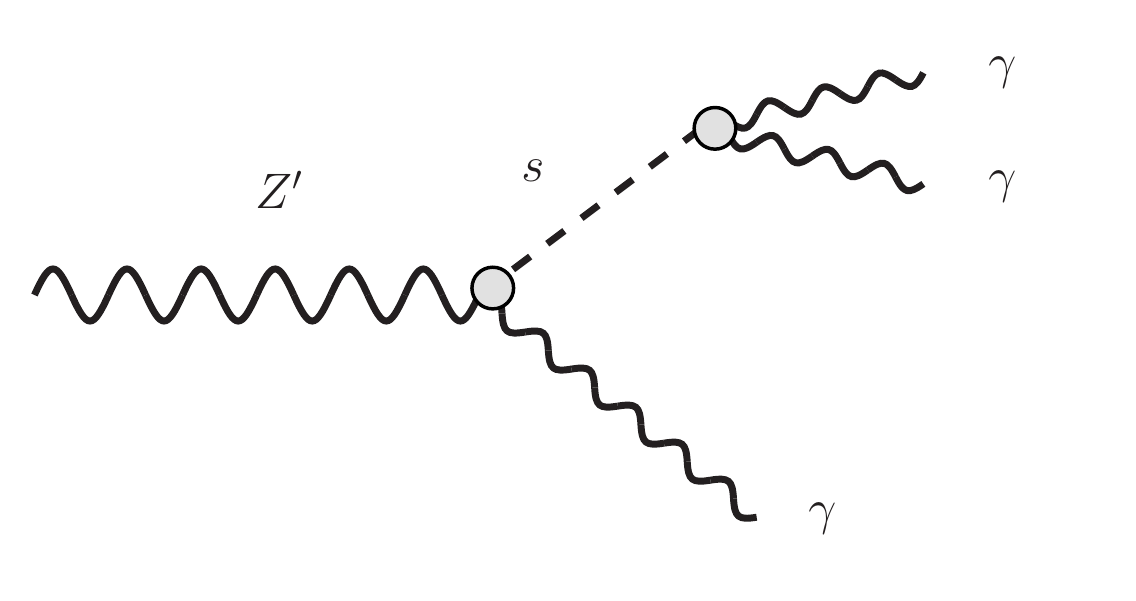}
\caption{Feynman diagram for the process $Z' \rightarrow 3 \gamma$}
\label{fig:feyn}
\end{figure}

The recently observed excess in the diphoton channel reported by ATLAS \cite{ATLAS-CONF-2015-081} and (with slightly less significance) CMS \cite{CMS-PAS-EXO-15-004}, has triggered a large amount of interest in the particle physics community \cite{Harigaya:2015ezk,Mambrini:2015wyu,Backovic:2015fnp,Angelescu:2015uiz,Buttazzo:2015txu,Knapen:2015dap,Nakai:2015ptz,Pilaftsis:2015ycr,Franceschini:2015kwy,DiChiara:2015vdm,Higaki:2015jag,McDermott:2015sck,Ellis:2015oso,Low:2015qep,Bellazzini:2015nxw,Gupta:2015zzs,Petersson:2015mkr,Molinaro:2015cwg,Dutta:2015wqh,Cao:2015pto,Matsuzaki:2015che,Kobakhidze:2015ldh,Martinez:2015kmn,Cox:2015ckc,Becirevic:2015fmu,No:2015bsn,Demidov:2015zqn,Chao:2015ttq,Fichet:2015vvy,Curtin:2015jcv,Bian:2015kjt,Chakrabortty:2015hff,Ahmed:2015uqt,Agrawal:2015dbf,Csaki:2015vek,Falkowski:2015swt,Aloni:2015mxa,Bai:2015nbs,Gabrielli:2015dhk,Benbrik:2015fyz,Kim:2015ron,Alves:2015jgx,Megias:2015ory,Carpenter:2015ucu,Bernon:2015abk,Chao:2015nsm}. Interpretations in terms of particle physics models have mainly assumed a scalar or tensor resonance
at 750~GeV, corresponding to the invariant mass observed in the diphoton system.
The case of a vector resonance seems problematic due to the Landau-Yang theorem~\cite{Yang:1950rg,Landau:1948kw}, which implies that a vector resonance cannot decay into two photons. 
In this letter we show that the observed excess can nevertheless be due to a new vector resonance, which decays into three photons (see also~\cite{Toro:2012sv}).
The idea is to have a vector which decays into a photon and a light singlet scalar
which then further decays into two photons; see figure~\ref{fig:feyn} for the Feynman diagram. If the scalar is light enough the two photons will be highly boosted and reconstructed as one
photon, leading to the apparent diphoton excess even though there are three photons in the final state.

A natural candidate for such a vector resonance is the gauge boson of a new $U(1)'$, which gets a mass from a hidden Higgs which is uncharged
under the Standard Model (SM) gauge group. Most $Z'$ models need additional fermions which have SM electroweak charges in order to be anomaly free and these additional fermions naturally induce the couplings which are needed in the current context.
The Higgs boson can naturally be much lighter than the $Z'$.

As an example we will consider the case of a new $U(1)'$ under which only the third generation quarks are charged.
This is motivated by the observation that no resonance has been observed at the LHC Run 1, implying that the increase of production
cross section needs to be sizeable. In the case of $b$ quarks in the initial state,
the ratio of production cross sections at 8 and 13 TeV is sufficiently large (5.4) due to the proton PDFs.
As couplings of $\mathcal{O}(1)$ are needed in order to have a large enough production cross section, the width
due to decays into $b\bar{b}$ and $t\bar{t}$ is sizeable and can easily be as large as $45$~GeV as indicated by the ATLAS result,
while being consistent with searches for $pp \rightarrow Z' \rightarrow \bar{b}b$ and $pp \rightarrow Z' \rightarrow \bar{t}t$.

This letter is organised as follows: 
In section~\ref{sec:eft} we introduce the effective approach and concentrate on the primary decay $Z' \rightarrow s \gamma$ while the subsequent decay $s\rightarrow \gamma\gamma$ is discussed in section~\ref{sec:2to1}. Details concerning possible ultraviolet (UV) completions are presented in section~\ref{sec:unitarity}. Finally, section~\ref{sec:discussion} discusses additional constraints and future prospects.

\section{Model-independent parametrisation and \texorpdfstring{$Z'$}{Z'} decays}
\label{sec:eft}

In the following we will study the case of a new $U(1)'$ gauge group under which only the third generation quarks are charged.
We assume the $U(1)'$ factor to be spontaneously broken by a complex scalar $S$ which is charged under the $U(1)'$ but singlet
with respect to the SM gauge group. 
The Lagrangian we consider is
\begin{align}
  \mathcal{L} = \quad & \mathcal{L}_\text{SM}  + \sum_{f = b,t} g' q_f Z'^\mu \, \bar{f} \gamma_\mu f  - \frac{1}{4} F'^{\mu\nu}F'_{\mu\nu} \nonumber \\ & + (D^\mu S)^\dagger D_\mu S + \mu_s^2 \, S^\dagger S - \lambda_s \left(S^\dagger S  \right)^2 ,
\end{align}
where $D^\mu=\partial^\mu + i \, g' q_S \, Z'^\mu$, $F'^{\mu\nu} = \partial^\mu Z'^\nu - \partial^\nu Z'^\mu$, and we have assumed vectorial couplings 
to quarks only\footnote{See the discussion in \cite{Kahlhoefer:2015bea} for why axial couplings can be problematic.}.
Once the hidden Higgs acquires a vev $w$ we define
\begin{equation}\label{eq:vev}
 S = \frac{1}{\sqrt{2}} (s + w)
\end{equation}
so that the Lagrangian in the unitary gauge can be written as
\begin{align}
 \mathcal{L} = \quad & \mathcal{L}_\text{SM} + \sum_{f = b,t} g' q_f Z'^\mu \, \bar{f} \gamma_\mu f  - \frac{1}{4} F'^{\mu\nu}F'_{\mu\nu} \nonumber \\ & + \frac{1}{2} m_{Z'}^2 \, Z'^\mu Z'_\mu  + \frac{1}{2}g'^2 q_S^2 \, Z'^\mu Z'_\mu (s^2 + 2\,s\,w) \nonumber \\ & + \frac{1}{2} \partial^\mu s \partial_\mu s
+ \frac{\mu_s^2}{2} (s+w)^2 - \frac{\lambda_s}{4} (s+w)^4 \; ,
\end{align}
with $m_{Z'}^2 = g'^2 q_S^2 w^2$. 
In addition, there can be mixing terms in the extended Higgs sector which we assume to be small for simplicity.
The model as written down above is anomalous and therefore additional fermions charged under the electroweak and $U(1)'$ gauge groups are required
to render the theory anomaly free. Interestingly, these additional fermions will generically induce effective couplings between the singlet $s$
and the two abelian gauge bosons, that we parametrise as
\begin{equation}
 \mathcal{L} \supset c_{B Z'} s B^{\mu\nu} F_{\mu\nu}' + c_{BB} s B^{\mu\nu} B_{\mu\nu} \;.
 \label{eq:EFT}
\end{equation}
As the anomaly cancellation conditions do not fix the additional field content uniquely, these coefficients are \textit{a priori} unknown. 
In the equation above, $B^{\mu\nu}$ stands for the $U(1)_Y$ field strength (note that there are hidden insertions of the vev of the hidden Higgs $S$ which is why these terms are gauge invariant).
In section~\ref{sec:unitarity} we will provide an estimate of the magnitude of these operators for a particular UV completion as an example.

The main production channel for the $Z'$ at the LHC is Drell-Yan production with $b$ quarks in the initial state.
As there is a large suppression due to the proton PDFs, this production cross section is typically rather small.
Nevertheless, for large couplings there are relevant constraints from resonance searches in $b\bar{b}$ and $t\bar{t}$ final states,
if the branching ratios are sizeable (which is naturally the case for the model under consideration).
Generally the partial widths of the $Z'$ into SM fermions can be written as
\begin{equation}
\Gamma(Z'\rightarrow f\bar{f}) = \frac{m_{Z'} N_c}{12\pi} 
 \sqrt{1-4\,x_f} \, (g' q_{f})^2 \left[1 + 2 x_f \right]  \; ,
\end{equation}
where $x_f = m_f^2/m_{Z'}^2$ and $N_c=3$ for quarks.
The decay modes into $b\bar{b}$ and $t\bar{t}$ will dominate the overall width. For $q_b = q_t$ the branching into each of the two channels is about 0.5. 
The partial widths of the $Z'$ to $s\gamma$ and $s Z$ are typically smaller and are given by
\begin{align}
 \Gamma(Z'\rightarrow s\gamma) = & \frac{(c_{B Z'})^2 c_W^2}{24\pi} m_{Z'}^3 \; , \\  \Gamma(Z'\rightarrow s Z) = & \frac{(c_{B Z'})^2 s_W^2}{24\pi} m_{Z'}^3 \; ,
\end{align}
where $c_W$ ($s_W$) stands for the cosine (sine) of the Weinberg angle.
Clearly the decay $Z' \rightarrow s Z$ is suppressed by a factor of $c_W^2/s_W^2\sim 3.3$ compared to $Z' \rightarrow s \gamma$. 
Finally the decay width of $s$ to $\gamma\gamma$ is given by
\begin{equation}
\Gamma(s\rightarrow\gamma\gamma) = \frac{(c_{BB})^2 c_W^4}{4\pi}m_s^3 \;.
\end{equation}
Several constraints on this model can be set in light of the results from the LHC Run~1.
The upper bounds on the production cross section times branching ratio into $b\bar{b}$ and $t\bar{t}$ final states are about 
1~pb~\cite{Khachatryan:2015tra} and 0.55~pb~\cite{Chatrchyan:2013lca} at the 95\% C.L., respectively. The diphoton channel is also constrained by Run 1 data to be below 2 fb at the 95\% C.L. \cite{Aad:2015mna}.
Given that the cross section of the observed excess at 13 TeV can be estimated to be around 5--10 fb, it is important
to have a significant increase of the production cross section with the centre of mass energy. For $b$ quarks in the initial state,
this is indeed the case with the ratio of cross sections at 8 and 13 TeV being slightly larger than 5.
Another constraint is that the total width of the resonance should not be too large (the best fit width measured by ATLAS
is about 45~GeV, although there is currently a large uncertainty on this number and CMS prefers a narrow width).
Potential constraints coming from electroweak precision data, in particular those due to the renormalization of the $Z$ vertices~\cite{deBlas:2015aea},
set negligible bounds as long as left-handed and right-handed quarks carry the same charge under the $U(1)'$ group.

\begin{figure}[t]
\centering
\includegraphics[width=0.95\columnwidth]{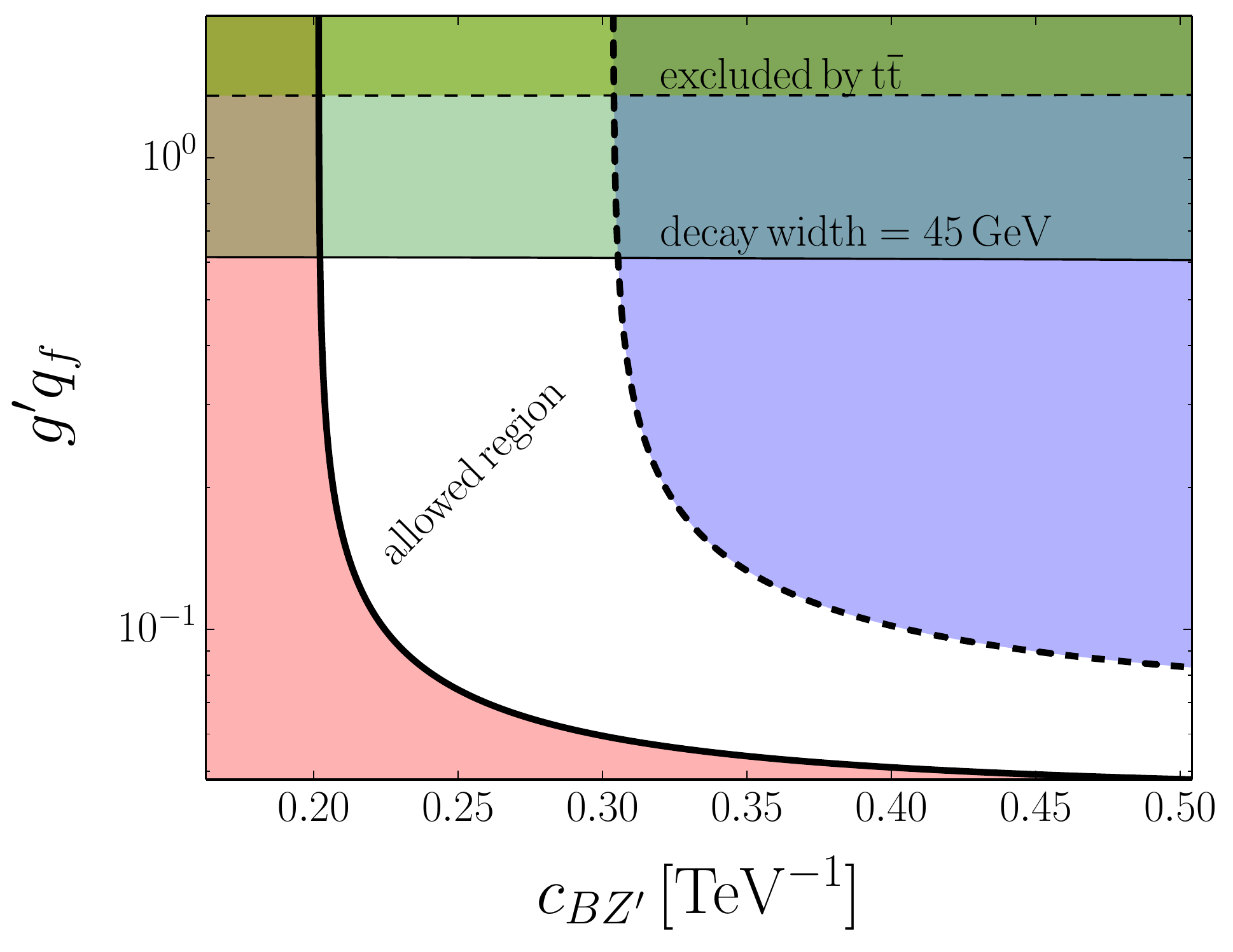}
\caption{Parameter space in the plane of the $Z'$ couplings to third generation quarks and to $s \gamma$ with bounds from LHC Run 1 diphotons (blue),
signal cross section below 5 fb (red) as well as bounds from $t\bar{t}$ resonance searches (green). The white region can explain the observed excess
and is consistent with all bounds.}
\label{fig:para}
\end{figure}

In order to study the LHC phenomenology we have implemented this generic parametrisation
including the five-dimensional operators in \texttt{MadGraph~5}~\cite{Alwall:2014hca} by means of \texttt{Feynrules v2}~\cite{Alloul:2013bka}.
We take $q_b = q_t \equiv q_f$ and fix the scalar mass $m_s=1$~GeV.\footnote{We show in the next section that the resulting two photons will be reconstructed as one.}
In figure~\ref{fig:para} we show the region in the $g'q_f-c_{BZ'}$ plane that can account for the
observed excess at 13 TeV without conflicting with previous searches. Indeed, the blue region
is excluded by diphoton searches at 8 TeV, while in the red region the cross section of the
diphoton signal at 13 TeV is smaller than 5 fb. Very large couplings to quarks produce too many
$t\bar{t}$ final states and are hence excluded. Searches for $b\bar{b}$ are less
constraining and not shown in the plot. Finally, a line indicating a
$Z'$ width of 45~GeV is also shown. Significantly larger widths are disfavoured as an explanation of the observed excess.

\section{Mimicking one photon with two}
\label{sec:2to1}

So far we have concentrated on the production cross section of the $Z'$ and its decay into $s \gamma$. Now we turn to the subsequent decays of the scalar $s$ in order to determine whether the observed excess can be reproduced in our model. The central observation is that the effective Lagrangian introduced in eq.~(\ref{eq:EFT}) does not only account for the production of $s$ from decays of the $Z'$, but it also leads to the decay processes $s \rightarrow \gamma\gamma$, $s \rightarrow ZZ$ and $s \rightarrow Z \gamma$. Provided the scalar is sufficiently light, only the first decay channel will be kinematically allowed,\footnote{We emphasise that any additional fermions should be heavier than about 100~GeV given their electroweak quantum numbers and that mixing in the Higgs sector is negligible. Consequently the branching fraction into fermions of the scalar $s$ is expected to be small.} so that we can assume $\text{BR}(s \rightarrow \gamma\gamma) = 1$.

Our central observation is that the decay chain $Z' \rightarrow s \gamma \rightarrow 3\gamma$ may be interpreted as a diphoton resonance, provided the two photons produced in the scalar decay are highly collimated (see for instance~\cite{Jaeckel:2015jla} for a similar analysis in the context of LEP). If the scalar $s$ is produced with a boost factor $\gamma_s$ and velocity $\beta_s$ in the laboratory frame, the distribution of opening angles $\alpha$ is given by
\begin{equation}
 \frac{\mathrm{d}N}{\mathrm{d}\alpha} = \frac{1}{4 \, \gamma_s \, \beta_s}\frac{\cos \alpha/2}{\sin^2 \alpha/2}\frac{1}{\sqrt{\gamma_s^2 \, \sin^2 \alpha/2 - 1}} \; ,
\end{equation}
which (assuming $\gamma_s \gg 1$) is strongly peaked towards the minimum opening angle $\alpha_\text{min} \simeq 2/\gamma_s$. Integrating the above distribution, we find that in 90\% of the scalar decays the opening angle between the two photons is smaller than $4.6 / \gamma_s$.

The typical angular resolution of the electromagnetic calorimeter in ATLAS and CMS is of order $20\:\text{mrad}$~\cite{ATLAS-CONF-2012-123,Chatrchyan:2013dga}. Hence, for $\gamma_s \gtrsim 200$, the vast majority of the scalar decays will appear in the detector as a single photon rather than two individual photons. This number agrees with an ATLAS study on the process $h \rightarrow aa \rightarrow 4\gamma$ for a light pseudoscalar $a$~\cite{ATLAS-CONF-2012-079}. It was found that for $m_a < 400\:\text{MeV}$, the two photons from the pseudoscalar decay are indistinguishable and the decay chain therefore mimics $h \rightarrow \gamma\gamma$ in the detector. Since $m_h = 125\:\text{GeV}$, this upper bound on the mass corresponds to the requirement $\gamma_a > 150$.\footnote{We thank Kerstin Tackmann and James Beacham for insightful discussions on this point.}

The consideration above leads us to consider a very light scalar with mass $m_s < 2\,\text{GeV}$. On the other hand, since we require the scalar to decay within the detector, it cannot be arbitrarily light. The decay length is given by
\begin{equation}
l = \beta_s \, \gamma_s / \Gamma  \approx \frac{4 \pi \, E_s}{c_{BB}^2 \, c_W^4 \, m_{s}^4} \;.
\end{equation}
Clearly, we must require $l \ll 1\:\text{m}$ in order for the majority of the scalars to decay within the detector, but it is certainly possible to have $l \sim 10\:\text{cm}$.\footnote{Note that it is not easily possible to reconstruct the tracks of the two photons and hence there is no problem with the scalar decaying from a displaced vertex (see also~\cite{Knapen:2015dap}).}

\begin{figure}[t]
\centering
\includegraphics[width=0.95\columnwidth]{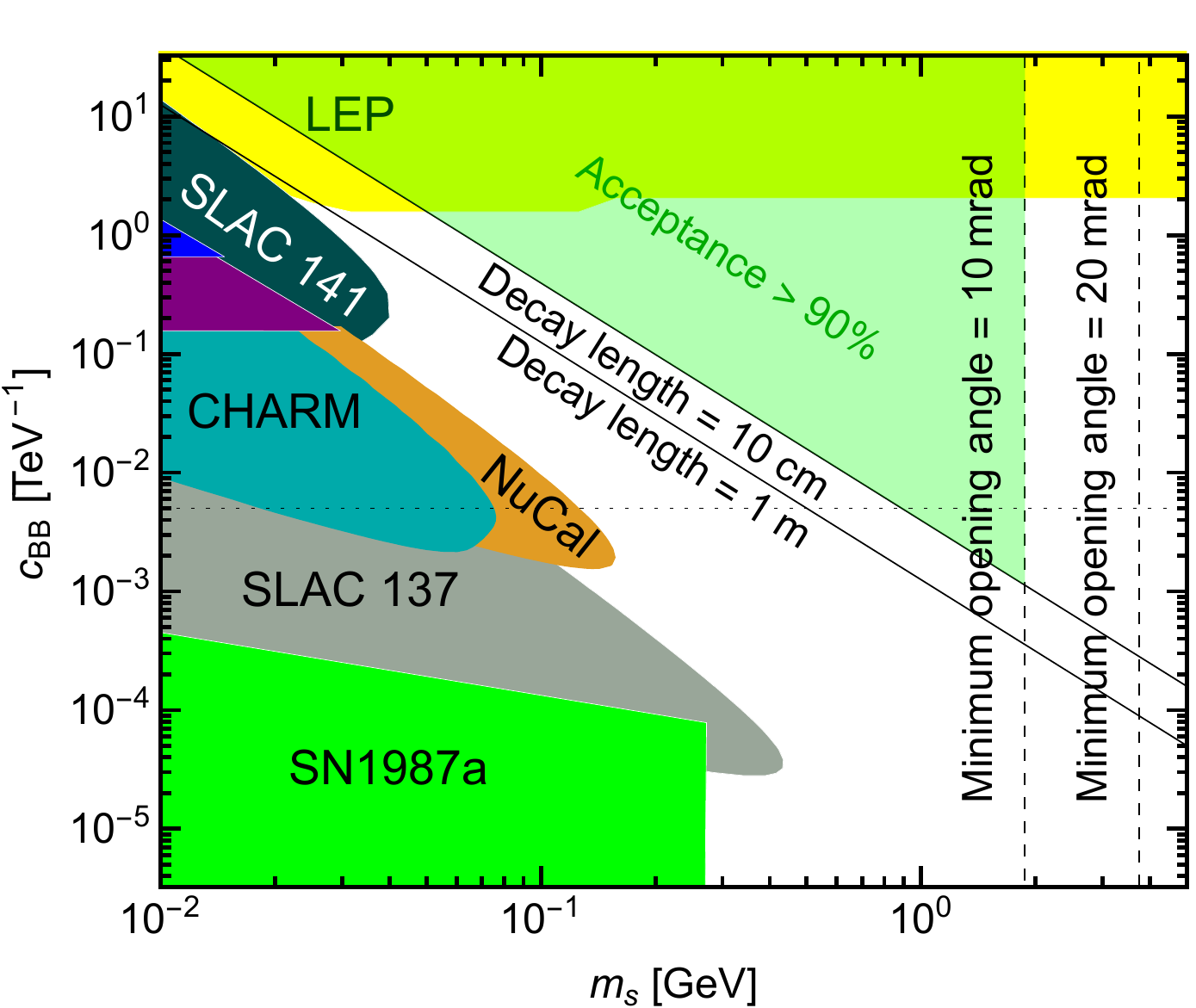}
\caption{Constraints and interesting regions in the $m_s-c_{BB}$ parameter plane. The solid (dashed) lines correspond to constant scalar decay length (photon opening angle). In the green shaded region we expect the acceptance to be larger than $90\%$, meaning that the large majority of scalars will decay within the detector and give a signal that mimics a single photon. For comparison, we also show bounds on the parameter space from beam-dump experiments and collider searches (see~\cite{Dobrich:2015jyk}).}
\label{fig:decay}
\end{figure}

In figure~\ref{fig:decay} we illustrate these considerations in the $m_s-c_{BB}$ parameter plane. Since we expect $c_{BB}$ to be comparable to $c_{BZ'}$, i.e.\ $c_{BB} \sim 10^{-1}\:\text{TeV}^{-1}$ we conclude that $m_s$ should not be smaller than about $200\:\text{MeV}$. In the example presented below, we find $c_{BB} \sim 5 \times 10^{-3} \: \text{TeV}^{-1}$ (indicated by the dotted horizontal line), which would limit $m_s$ to a relatively narrow range around $1\text{--}2\:\text{GeV}$. In this mass range it is possible to have a sufficiently short decay length and at the same time a sufficiently small opening angle between the two photons produced in the decay.

\section{A \texorpdfstring{$U(1)'$}{U(1)'} model example}
\label{sec:unitarity}

\begin{table*}[t]
\begin{center}
\begin{tabular}{ccccc}
\hline \hline
~~Field~~  & ~~$SU(3)_C$~~ & ~~$SU(2)_L$~~ & ~~$U(1)_Y$~~ & ~~$U(1)^\prime$~~ \\
\hline \hline
 $\Sigma_L$   & $\mathbf{1}$  & $\mathbf{2}$ & $Y_1$ & $Q_1$ \\
 $\Sigma_R$   & $\mathbf{1}$  & $\mathbf{2}$ & $Y_1$ & $Q_2$ \\
 $\rho_R$   & $\mathbf{1}$  & $\mathbf{1}$ & $Y_2$ & $Q_1$ \\
 $\rho_L$   & $\mathbf{1}$  & $\mathbf{1}$ & $Y_2$ & $Q_2$ \\
 $\xi_R$   & $\mathbf{1}$  & $\mathbf{1}$ & $Y_3$ & $Q_1$ \\
 $\xi_L$   & $\mathbf{1}$  & $\mathbf{1}$ & $Y_3$ & $Q_2$ \\
\hline
 $S$        & $\mathbf{1}$  & $\mathbf{1}$ & $0$   & $Q_1 - Q_2 = -3  q_f$ \\
\hline\hline
\end{tabular}
\caption{Fermionic and scalar field content of a possible anomaly-free UV completion, and the corresponding quantum numbers~\cite{Duerr:2013dza}.\label{tab:newParticles}}
\end{center}
\end{table*}

Having discussed the excess in a model-independent framework up to now, let us now give an example for a UV complete model that has the required properties. 
For our analysis we have assumed that only the third generation quarks are charged under the new $U(1)'$ gauge group, with $q_t = q_b \equiv q_f$. As commented earlier, new fermionic states charged under the $U(1)^\prime$ and the SM gauge group are required to make this model anomaly free. It can easily be checked that all relevant anomalies are cancelled with the fermions given in Table~\ref{tab:newParticles}, which are vector-like under the SM gauge group and do not need to carry colour charge. This solution for the cancellation of anomalies was first found in the context of gauging baryon and lepton numbers~\cite{Duerr:2013dza}. The only requirements on the hypercharges and the $U(1)^\prime$ charges are
\begin{equation}\label{eq:conditionQ}
 Q_1 - Q_2 = -3  q_f
\end{equation}
and\footnote{Note that the linear condition on the hypercharges is only relevant if $|Q_1| \neq |Q_2|$.} 
\begin{align}
 2 Y_1^2 - Y_2^2 - Y_3^2 &= -\frac{1}{2},\\
 2 Y_1 - Y_2 - Y_3 &= 0 .
\end{align}
The latter conditions on the hypercharges are solved by choosing (see e.g.~\cite{Chao:2015nsm})
\begin{equation}
 \left( Y_1, Y_2, Y_3 \right) = (a, \ a+\frac{1}{2}, \ a-\frac{1}{2}) \; ,
\end{equation}
where $a$ is a free parameter. Due to $U(1)'$ gauge symmetry the new fermions cannot have bare mass terms, but they will have Yukawa interactions with the new scalar $S$ provided the $U(1)^\prime$ charge of the scalar $S$ satisfies the condition $q_S = Q_1 - Q_2$, which implies $q_S = -3 q_f$ from anomaly cancellation. Once $S$ obtains its vacuum expectation value, it will generate mass terms for the new fermions, as well as a mass term for the $Z^\prime$ gauge boson.

Since the additional fermions are charged under both $U(1)_Y$ and $U(1)'$, they will naturally generate the couplings $c_{BZ'}$ and $c_{BB}$ introduced in eq.~(\ref{eq:EFT}). As long as the additional fermions are heavy compared to $m_{Z^\prime}$ and $m_s$, we can estimate their contribution to be
\begin{align}
c_{BZ'} & = \mathop{\sum_\text{heavy}}_\text{fermions} \frac{g_Y \, g' \, Q_Y \, Q'}{12 \, \pi^2 \, w} \; , \nonumber \\
c_{BB} & = \mathop{\sum_\text{heavy}}_\text{fermions} \frac{g_Y^2 \, Q_Y^2}{24 \, \pi^2 \, w} \; ,
\end{align}
where $g_Y$ and $g'$ denote the gauge couplings of $U(1)_Y$ and $U(1)'$, respectively, $Q_Y$ and $Q'$ are the corresponding vector charges of the new fermions (i.e.\ $Q' = (Q'_L + Q'_R)/2$) and $w$ is the vev of the second Higgs as defined in eq.~(\ref{eq:vev}).

Substituting the $Z'$ mass for $w$, these expressions can be rewritten as
\begin{align}
c_{BZ'} & = \mathop{\sum_\text{heavy}}_\text{fermions} \frac{g_Y \, g'^2 \, Q_Y \, Q' \, q_f}{4 \, \pi^2 \, m_{Z'}} \; , \nonumber \\
c_{BB} & = \mathop{\sum_\text{heavy}}_\text{fermions} \frac{g_Y^2 \, Q_Y^2 \, g' \, q_f}{8 \, \pi^2 \, m_{Z'}} 
\end{align}
leading to
\begin{align}
c_{BZ'} & \sim 0.012\:\text{TeV}^{-1} \times \mathop{\sum_\text{heavy}}_\text{fermions} g'^2 \, Q_Y \, Q'\, q_f  \; , \nonumber \\ c_{BB} & \sim 0.002\:\text{TeV}^{-1} \times \mathop{\sum_\text{heavy}}_\text{fermions} Q_Y^2 \, g'\, q_f  
\end{align}
for $m_{Z'}=750$~GeV.
Since several fermions are expected to give a sizeable contribution to this expression, it is perfectly conceivable to obtain large enough effective couplings to account for the observed excess for gauge couplings and charges of order unity. 
For example, for $q_f = 1/3$ the bound on the total width of the $Z'$ implies $g' \leq 1.8$ (see figure~\ref{fig:para}). Assuming that this bound is saturated and substituting the charges from table~\ref{tab:newParticles}, we obtain
\begin{align}
c_{BZ'} & \sim 0.025\:\text{TeV}^{-1} \times a (Q_1 + Q_2)\; , \nonumber \\ c_{BB} & \sim 0.0012 \:\text{TeV}^{-1} \times \left(4 a^2 + \frac{1}{2}\right) \; .
\end{align}
Setting for example $a = 1$, $Q_1 = 4$ and $Q_2 = Q_1 + 3 \, q_f = 5$ gives $c_{BZ'} = 0.23 \:\text{TeV}^{-1}$ and $c_{BB} = 0.005 \:\text{TeV}^{-1}$, which are both consistent with the constraints shown in figures~\ref{fig:para} and~\ref{fig:decay}.
We note that rather large fermion charges are required in order to obtain sufficiently large couplings. However, the contributions of the new fermions to $c_{BZ'}$ and $c_{BB}$ will be larger if their masses are comparable to $m_{Z'}/2$, which is easily possible given that the new fermions do not carry colour charge. In such a set-up, smaller charges may be sufficient in order to explain the diphoton excess. A detailed study of the implication of such a UV completion is left to future work.

\section{Discussion}
\label{sec:discussion}

We have shown that extending the SM with a new $U(1)'$ gauge group can provide an explanation for the recently observed diphoton excess around $\sim 750$~GeV by both ATLAS and CMS. The central idea is that a very light scalar decaying into two photons can mimic a single photon if it is produced with sufficiently large boost factor. This observation allows us to evade the Landau-Yang theorem and to obtain the diphoton excess from a vector resonance.

As a specific example we have discussed the addition of a new complex scalar degree of freedom and a set of new fermions in order for the new gauge $Z'$ boson to have a mass and the full theory to remain anomaly free. It turns out that loops of new extra fermions give rise to trilinear vertices allowing the $Z'$ to decay into $s\gamma$ and the light scalar $s$ to decay into $\gamma \gamma$ respectively. The corresponding couplings are expected to be of the same order, namely $c_{BZ'} \sim c_{BB} \sim 0.1\:\text{TeV}^{-1}$, if the charges carried by the new fermions are large enough. Interestingly, these values can make the cross section into diphotons large enough to account for the observed excess and, at the same time, make the light scalar mimic a one-photon signature while still decaying within the detector.

In fact, the idea of reconstructing two photons as a single one may also be of interest to other models. One interesting example is provided by a scalar resonance $\Phi$ with mass $m_\Phi \sim 750$~GeV that decays into two very light scalars $s$ that subsequently decay into two photons. The relevant interactions can be written as
\begin{equation}
\mathcal{L} \supset c_{GG}\,\Phi\, G_{\mu\nu}^aG^{\mu\nu\,a} + \kappa_\Phi \Phi s^2.
\end{equation}
\begin{figure}[t]
\centering
\includegraphics[width=0.95\columnwidth]{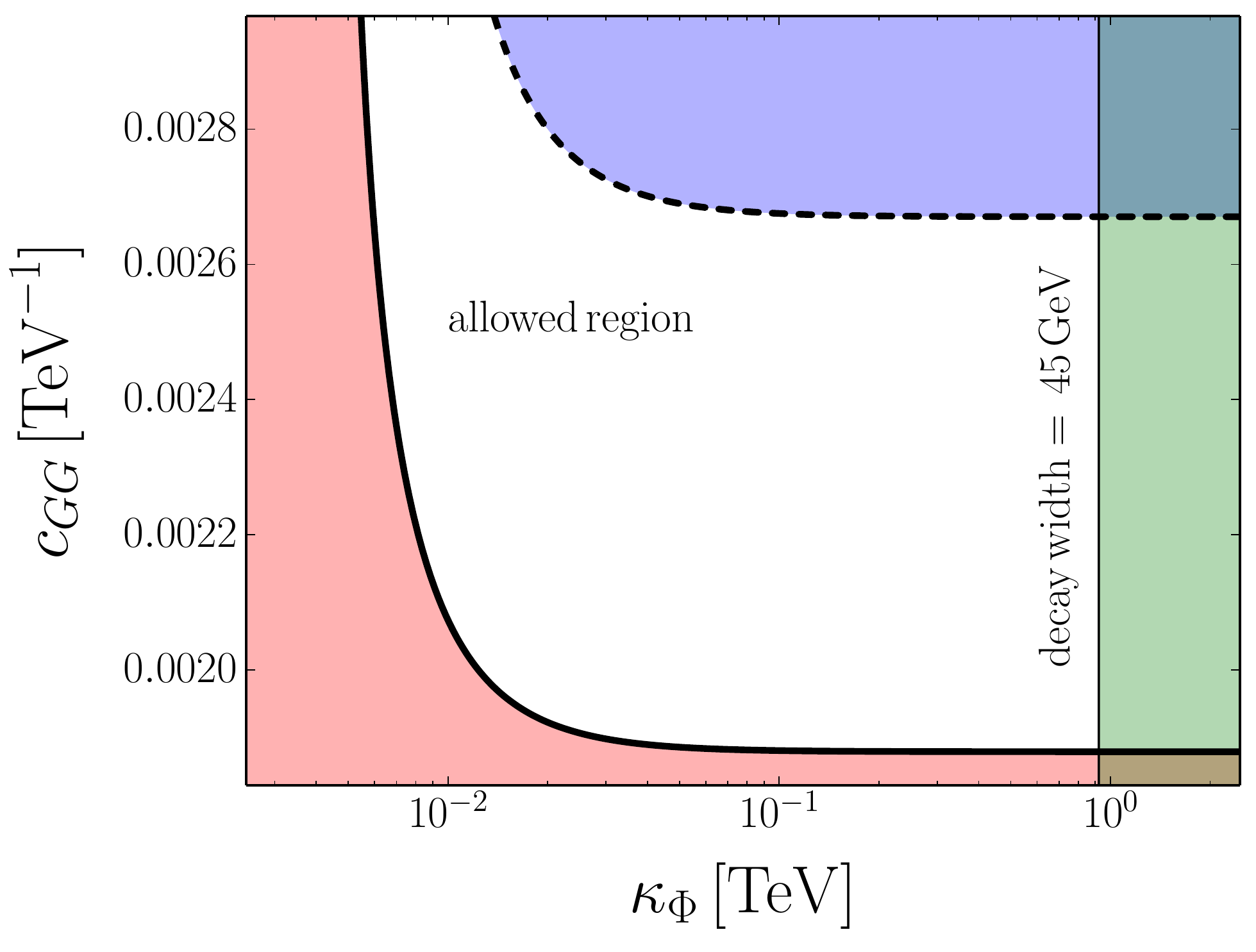}
\caption{Allowed and excluded parameter space regions in the plane of the $\Phi$ couplings to gluons and to $s$. }
\label{fig:spara}
\end{figure}
In figure~\ref{fig:spara} we show the parameter space region in the $c_{GG}-\kappa_{\Phi}$ plane allowed by the diphoton searches at both 8 and 13 TeV as well as the best fit width. Searches for dijets are not sensitive to this region of the parameter space.

The two scenarios can be easily distinguished with additional data. If the diphoton excess is due to the process $Z' \rightarrow s \gamma$, a corresponding signal in $Z' \rightarrow s Z \rightarrow \gamma\gamma Z$ should be discovered in the near future. On the other hand, no such additional decay channel is expected if the diphoton excess is due to a scalar decaying into two very light scalars.

Finally, we emphasize that the gauge boson of a new $U(1)'$ gauge group can mix with the neutral gauge bosons of the SM. Indeed, given that we include fermions charged under both $U(1)_Y$ and $U(1)'$, loop processes will in general lead to kinetic mixing of the form $- \frac{1}{2} \sin \epsilon F'^{\mu\nu} B_{\mu\nu}$~\cite{Holdom:1985ag,Carone:1995pu}. The presence of such a kinetic mixing term can modify electroweak precision observables and lead to sizeable decays of the $Z'$ into SM leptons, providing promising ways to constrain this scenario. However, the precise magnitude of these effects depends on the specific UV completion. We leave a detailed investigation of these effects to future studies.

\section*{Acknowledgments}

We thank James Beacham and Kerstin Tackmann for very helpful discussions and Joerg Jaeckel for valuable comments on the draft. This work is supported by the German Science Foundation (DFG) under the Collaborative Research Center (SFB) 676 Particles, Strings and the Early Universe as well as the ERC Starting Grant `NewAve' (638528).


\section*{References}

\begin{thebibliography}{10}

\bibitem{ATLAS-CONF-2015-081}
ATLAS Collaboration,
\newblock CERN preprint ATLAS-CONF-2015-081 (2015).

\bibitem{CMS-PAS-EXO-15-004}
CMS Collaboration,
\newblock CERN preprint CMS-PAS-EXO-15-004 (2015).

\bibitem{Harigaya:2015ezk}
K. Harigaya and Y. Nomura,
\newblock Phys. Lett. B754 (2016) 151-156, 1512.04850.

\bibitem{Mambrini:2015wyu}
Y. Mambrini, G. Arcadi and A. Djouadi,
\newblock (2015), 1512.04913.

\bibitem{Backovic:2015fnp}
M. Backovic, A. Mariotti and D. Redigolo,
\newblock (2015), 1512.04917.

\bibitem{Angelescu:2015uiz}
A. Angelescu, A. Djouadi and G. Moreau,
\newblock (2015), 1512.04921.

\bibitem{Buttazzo:2015txu}
D. Buttazzo, A. Greljo and D. Marzocca,
\newblock (2015), 1512.04929.

\bibitem{Knapen:2015dap}
S. Knapen et~al.,
\newblock (2015), 1512.04928.

\bibitem{Nakai:2015ptz}
Y. Nakai, R. Sato and K. Tobioka,
\newblock (2015), 1512.04924.

\bibitem{Pilaftsis:2015ycr}
A. Pilaftsis,
\newblock (2015), 1512.04931.

\bibitem{Franceschini:2015kwy}
R. Franceschini et~al.,
\newblock (2015), 1512.04933.

\bibitem{DiChiara:2015vdm}
S. Di~Chiara, L. Marzola and M. Raidal,
\newblock (2015), 1512.04939.

\bibitem{Higaki:2015jag}
T. Higaki et~al.,
\newblock (2015), 1512.05295.

\bibitem{McDermott:2015sck}
S.D. McDermott, P. Meade and H. Ramani,
\newblock (2015), 1512.05326.

\bibitem{Ellis:2015oso}
J. Ellis et~al.,
\newblock (2015), 1512.05327.

\bibitem{Low:2015qep}
M. Low, A. Tesi and L.T. Wang,
\newblock (2015), 1512.05328.

\bibitem{Bellazzini:2015nxw}
B. Bellazzini et~al.,
\newblock (2015), 1512.05330.

\bibitem{Gupta:2015zzs}
R.S. Gupta et~al.,
\newblock (2015), 1512.05332.

\bibitem{Petersson:2015mkr}
C. Petersson and R. Torre,
\newblock (2015), 1512.05333.

\bibitem{Molinaro:2015cwg}
E. Molinaro, F. Sannino and N. Vignaroli,
\newblock (2015), 1512.05334.

\bibitem{Dutta:2015wqh}
B. Dutta et~al.,
\newblock (2015), 1512.05439.

\bibitem{Cao:2015pto}
Q.H. Cao et~al.,
\newblock (2015), 1512.05542.

\bibitem{Matsuzaki:2015che}
S. Matsuzaki and K. Yamawaki,
\newblock (2015), 1512.05564.

\bibitem{Kobakhidze:2015ldh}
A. Kobakhidze et~al.,
\newblock (2015), 1512.05585.

\bibitem{Martinez:2015kmn}
R. Martinez, F. Ochoa and C.F. Sierra,
\newblock (2015), 1512.05617.

\bibitem{Cox:2015ckc}
P. Cox et~al.,
\newblock (2015), 1512.05618.

\bibitem{Becirevic:2015fmu}
D. Becirevic et~al.,
\newblock (2015), 1512.05623.

\bibitem{No:2015bsn}
J.M. No, V. Sanz and J. Setford,
\newblock (2015), 1512.05700.

\bibitem{Demidov:2015zqn}
S.V. Demidov and D.S. Gorbunov,
\newblock (2015), 1512.05723.

\bibitem{Chao:2015ttq}
W. Chao, R. Huo and J.H. Yu,
\newblock (2015), 1512.05738.

\bibitem{Fichet:2015vvy}
S. Fichet, G. von Gersdorff and C. Royon,
\newblock (2015), 1512.05751.

\bibitem{Curtin:2015jcv}
D. Curtin and C.B. Verhaaren,
\newblock (2015), 1512.05753.

\bibitem{Bian:2015kjt}
L. Bian et~al.,
\newblock (2015), 1512.05759.

\bibitem{Chakrabortty:2015hff}
J. Chakrabortty et~al.,
\newblock (2015), 1512.05767.

\bibitem{Ahmed:2015uqt}
A. Ahmed et~al.,
\newblock (2015), 1512.05771.

\bibitem{Agrawal:2015dbf}
P. Agrawal et~al.,
\newblock (2015), 1512.05775.

\bibitem{Csaki:2015vek}
C. Csaki, J. Hubisz and J. Terning,
\newblock (2015), 1512.05776.

\bibitem{Falkowski:2015swt}
A. Falkowski, O. Slone and T. Volansky,
\newblock (2015), 1512.05777.

\bibitem{Aloni:2015mxa}
D. Aloni et~al.,
\newblock (2015), 1512.05778.

\bibitem{Bai:2015nbs}
Y. Bai, J. Berger and R. Lu,
\newblock (2015), 1512.05779.

\bibitem{Gabrielli:2015dhk}
E. Gabrielli et~al.,
\newblock (2015), 1512.05961.

\bibitem{Benbrik:2015fyz}
R. Benbrik, C.H. Chen and T. Nomura,
\newblock (2015), 1512.06028.

\bibitem{Kim:2015ron}
J.S. Kim et~al.,
\newblock (2015), 1512.06083.

\bibitem{Alves:2015jgx}
A. Alves, A.G. Dias and K. Sinha,
\newblock (2015), 1512.06091.

\bibitem{Megias:2015ory}
E. Megias, O. Pujolas and M. Quiros,
\newblock (2015), 1512.06106.

\bibitem{Carpenter:2015ucu}
L.M. Carpenter, R. Colburn and J. Goodman,
\newblock (2015), 1512.06107.

\bibitem{Bernon:2015abk}
J. Bernon and C. Smith,
\newblock (2015), 1512.06113.

\bibitem{Chao:2015nsm}
W. Chao,
\newblock (2015), 1512.06297.

\bibitem{Yang:1950rg}
C.N. Yang,
\newblock Phys. Rev. 77 (1950) 242.

\bibitem{Landau:1948kw}
L.D. Landau,
\newblock Dokl. Akad. Nauk Ser. Fiz. 60 (1948) 207.

\bibitem{Toro:2012sv}
N. Toro and I. Yavin,
\newblock Phys. Rev. D86 (2012) 055005, 1202.6377.

\bibitem{Kahlhoefer:2015bea}
F. Kahlhoefer et~al.,
\newblock (2015), 1510.02110.

\bibitem{Khachatryan:2015tra}
CMS Collaboration, V. Khachatryan et~al.,
\newblock JHEP 11 (2015) 071, 1506.08329.

\bibitem{Chatrchyan:2013lca}
CMS Collaboration, S. Chatrchyan et~al.,
\newblock Phys. Rev. Lett. 111 (2013) 211804, 1309.2030,
\newblock [Erratum: Phys. Rev. Lett.112,no.11,119903(2014)].

\bibitem{Aad:2015mna}
ATLAS Collaboration, G. Aad et~al.,
\newblock Phys. Rev. D92 (2015) 032004, 1504.05511.

\bibitem{deBlas:2015aea}
J. de~Blas, M. Chala and J. Santiago,
\newblock JHEP 09 (2015) 189, 1507.00757.

\bibitem{Alwall:2014hca}
J. Alwall et~al.,
\newblock JHEP 07 (2014) 079, 1405.0301.

\bibitem{Alloul:2013bka}
A. Alloul et~al.,
\newblock Comput. Phys. Commun. 185 (2014) 2250, 1310.1921.

\bibitem{Jaeckel:2015jla}
J. Jaeckel and M. Spannowsky,
\newblock (2015), 1509.00476.

\bibitem{ATLAS-CONF-2012-123}
ATLAS Collaboration,
\newblock CERN preprint ATLAS-CONF-2012-123 (2012).

\bibitem{Chatrchyan:2013dga}
CMS Collaboration, S. Chatrchyan et~al.,
\newblock JINST 8 (2013) P09009, 1306.2016.

\bibitem{ATLAS-CONF-2012-079}
ATLAS Collaboration,
\newblock CERN preprint ATLAS-CONF-2012-079 (2012).

\bibitem{Dobrich:2015jyk}
B. Dobrich et~al.,
\newblock (2015), 1512.03069.

\bibitem{Duerr:2013dza}
M. Duerr, P. Fileviez~Perez and M.B. Wise,
\newblock Phys. Rev. Lett. 110 (2013) 231801, 1304.0576.

\bibitem{Holdom:1985ag}
B. Holdom,
\newblock Phys. Lett. B166 (1986) 196.

\bibitem{Carone:1995pu}
C.D. Carone and H. Murayama,
\newblock Phys. Rev. D52 (1995) 484, hep-ph/9501220.

\end{thebibliography}
\end{document}